\documentclass[preprint,12pt]{elsarticle}

\usepackage{amssymb}
\usepackage{algorithm}
\usepackage{algorithmic}
\usepackage{multirow}
\usepackage{adjustbox}
\usepackage{graphicx}
\usepackage{subcaption}
\usepackage{booktabs}

\usepackage{array}
\usepackage[english]{babel}
\usepackage[utf8x]{inputenc}
\usepackage[colorinlistoftodos]{todonotes}
\usepackage{xcolor}
\usepackage{soul}

\usepackage{lineno}

\journal{In the Cloud, Fog and Mist Computing - Resource Allocation and Scheduling Perspectives}

\begin{document}


\begin{frontmatter}

\title{Empirical Investigation of Factors influencing Function as a Service Performance in Different Cloud/Edge System Setups}

\author{Anastasia-Dimitra Lipitakis}

\affiliation{
            organization={Department of Informatics and Telematics},
            addressline={Harokopio University of Athens}, 
            city={Athens},
            postcode={GR 177 78}, 
            country={Greece}}
\author{George Kousiouris}

\affiliation{organization={Department of Informatics and Telematics},
            addressline={Harokopio University of Athens}, 
            city={Athens},
            postcode={GR 177 78}, 
            country={Greece}}

\author{Mara Nikolaidou}

\affiliation{organization={Department of Informatics and Telematics},
            addressline={Harokopio University of Athens}, 
            city={Athens},
            postcode={GR 177 78}, 
            country={Greece}}

\author{Cleopatra Bardaki}

\affiliation{organization={Department of Informatics and Telematics},
            addressline={Harokopio University of Athens}, 
            city={Athens},
            postcode={GR 177 78}, 
            country={Greece}}

\author{Dimosthenis Anagnostopoulos}

\affiliation{organization={Department of Informatics and Telematics},
            addressline={Harokopio University of Athens}, 
            city={Athens},
            postcode={GR 177 78}, 
            country={Greece}}

\begin{abstract}
Experimental data can aid in gaining insights about a system operation, as well as determining critical aspects of a modelling or simulation process. In this paper, we analyze the data acquired from an extensive experimentation process in a serverless Function as a Service system (based on the open source Apache Openwhisk) that has been deployed across 3 available cloud/edge locations with different system setups. Thus, they can be used to model distribution of functions through multi-location aware scheduling mechanisms. The experiments include different traffic arrival rates, different setups for the FaaS system, as well as different configurations for the hardware and platform used. We analyse the acquired data for the three FaaS system setups  and discuss their differences presenting interesting conclusions with relation to transient effects of the system, such as the effect on wait and execution time. We also demonstrate interesting trade-offs with relation to system setup and indicate a number of factors that can affect system performance and should be taken under consideration in modelling attempts of such systems. 
\end{abstract}

\begin{graphicalabstract}

\end{graphicalabstract}


\begin{highlights}
\item Experimental data analysis of 3 different setups of a FaaS system.
\item Insights into the transient effects of the system, as well as considerations in terms of modelling approaches
\item Explore interesting conclusions regarding the system setup and its trade-offs, as well as hidden factors influencing performance.
\end{highlights}


\end{frontmatter}


\section{Introduction}
\label{intro}
\noindent Serverless computing \cite{jonas2019cloud} is a cloud computing model that enables developers to build and run applications without worrying about the underlying infrastructure. This architecture breaks down applications into small, stateless functions that can be executed in response to events \cite{majewski2021algorithms}, such as HTTP requests, database changes, messaging notifications or other means of triggering. Function-as-a-Service (FaaS) systems are the key enabler for serverless computing, providing an environment in which developers can deploy, manage, and execute these functions. \par
\begin{figure*}[htbp]
\includegraphics[width=\textwidth]{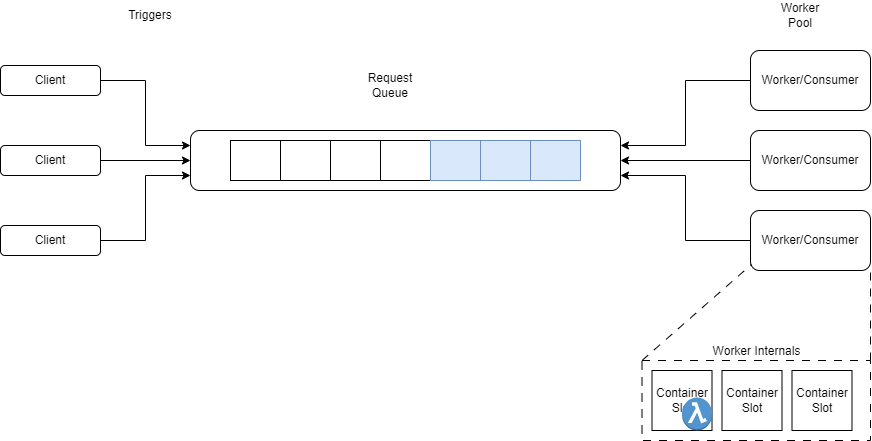}
\caption{Overview of a FaaS System Architecture}
\label{fig:Figure1}
\end{figure*}

\noindent FaaS systems, like the open source Apache Openwhisk \cite{openwhisk}, typically rely on the queue based load levelling pattern with competing consumers \cite{patterns} (Fig. \ref{fig:Figure1}). Incoming requests are queued so that they do not create congestion on the back-end, while worker nodes consume them (at their own pace) in a First Come First Serve (FCFS) basis. Through this architecture, back-ends are relieved from bursty traffic and automatic load balancing is achieved, while scalability can be regulated through adding or removing worker instances. \par

\noindent This generic FaaS architecture introduces a system uniformity, since all registered functions follow the same execution model. Thus, generic modelling mechanisms, such as queuing models\cite{mahmoudi2019optimizing},  can be used for predicting a system's performance with relation to factors, such as response time, wait time in the system, concurrent clients in the system etc., while needing generic inputs that can be easily obtained (e.g. incoming request rate, function duration etc). Thus, they alleviate from the burden of gathering sufficiently large and diverse datasets needed for training black box methods like ANNs, while they can easily produce simulated results to help drive decision making (e.g. on the size of the used cluster). \par
\noindent In this work, the goal is to analyze and get insights from the data initially acquired from an extensive experimentation process in \cite{ehealth} against 3 different types of FaaS systems. The aim of the process is to calculate example needed parameters for modeling approaches, validating potential assumptions required by different modelling methodologies as well as highlight some esting phenomena that occur during the specific system operation.\par
\noindent Furthermore, the knowledge of how each different location performs in a multi-cloud-edge setup can aid higher level scheduling mechanisms to optimally distribute incoming requests based on needed constraints. Effective resource allocation and scheduling are essential to ensure optimal performance and cost-effectiveness in such environments \cite{gadepalli2019challenges,kijak2018challenges,fang2020efficient,luo2021resource}, balancing between local resource contention in small size clusters and remote, latency-increasing execution in large ones. \par

\noindent A preliminary, high level investigation of the acquired dataset is presented in \cite{ehealth} mainly from an average times point of view. In this paper, we further analyze the acquired data, include the transitioned time aspect and further evaluate the collected data.\par

\noindent The paper is organized as follows: In Section 2, related work is portrayed with relation to FaaS system performance. In Section 3, the experimental setup, as well as the raw data graphs are presented, while in section 4 they are further investigated and key take-aways are extracted from them. Conclusions reside in Section 5.\par

\section{Related Work}
\label{relatedwork}

\noindent The adoption of cloud and edge computing has witnessed a significant surge, with Function as a Service (FaaS) emerging as a prominent paradigm for executing serverless computing workloads. As FaaS systems become increasingly prevalent and diverse, there is a growing need to understand their performance characteristics and optimize their operation across various cloud and edge environments. This section provides an overview of the existing literature related to the performance investigation of cloud/edge FaaS systems and the extraction of baseline data.\par

\noindent Evaluating the performance of Function as a Service (FaaS) systems deployed in cloud environments is a crucial research topic that sheds light on the benefits and challenges of this emerging cloud computing paradigm. Numerous studies have focused on measuring and analyzing various performance aspects of FaaS platforms, including latency, throughput, scalability, resource utilization, and cost. Seneviratne et al. \cite{seneviratne2023taxonomy} discuss the performance challenges of FaaS and serverless architectures, identifying four main research directions: benchmarking, modeling, optimization, and testing. Scheuner et al. \cite{scheuner2020function} provide a comprehensive overview of existing empirical studies on FaaS performance, synthesizing their findings and implications. Tan et al.\cite{tan2021low} propose a model that depicts the cold-start performance of FaaS systems and introduce SnapFaaS, a snapshot design that achieves near-optimal cold-start performance. Additionally, Fard et al. [18] present a comprehensive survey of existing resource allocation techniques in cloud computing, classifying them based on criteria such as resource type, allocation objective, allocation strategy, and allocation algorithm. In addition to evaluating FaaS systems, performance evaluation studies have also explored the analysis of log files in cloud computing environments \cite{mavridis2017performance}. \par
\noindent In parallel, the adoption of edge computing, which aims to bring computation closer to data sources and consumers, has gained significant attention due to its potential to mitigate latency and reduce bandwidth consumption. Within this context, the integration of FaaS into edge computing enables the realization of serverless edge computing (SEC), facilitating the development and deployment of event-driven net of Things (IoT) applications.\par
\noindent The effective optimization of SEC poses various challenges, including data transfer, function placement, resource allocation, and scheduling. To address these challenges, several studies have proposed diverse solutions. 
Cicconetti et al. \cite{cicconetti2022faas} compare three execution models for stateful workflows within edge networks: pure FaaS, StateProp (involving propagation of the application state throughout the entire chain of functions), and StateLocal (with localized state storage at the workers executing the functions). The authors demonstrate that applying the principle of data locality can effectively reduce network traffic and enhance end-to-end delay performance. Furthermore, Yao et al. \cite{yao2023performance} introduces a performance optimization framework for SEC function placement, leveraging reinforcement learning and graph neural networks. This framework dynamically adjusts the function placement based on prevailing network conditions and specific application requirements.\par
\noindent With relation to other available open datasets, two cases that are frequently used are the ones from \cite{shahrad2020serverless} and \cite{zhang2021faster}. However, these traces primarily depict function invocation workloads and frequencies and are not linked to the inner resource setup or utilization. Furthermore, they include only the function execution duration, which is only one part of the total latency experienced by the end user.
In our case, the delays include also network latency, wait time in the system, as well as execution time for different system setups.  By analyzing the acquired data, we aim to uncover transient effects of the system, such as the effect on wait and execution time, concurrent container overheads and identify esting trade-offs in system setup that can inform decision-making for optimization.\par

\section{Experimental Setup and Data Presentation}
\label{setup}

\subsection{Experiment Setup Description }
\noindent Experimental data plays a crucial role in gaining insights into system operations and identifying critical aspects of modeling or simulation processes. In this paper, the presented data are acquired from an extensive experimentation process across three diverse cloud/edge locations (Harokopio University, AWS Sweden and Azure Netherlands), each with unique characteristics. In each case, one VM node was used (with different characteristics in terms of CPU and memory) inside which Openwhisk was deployed. The function invocation rates employed were 12, 30, and 60 messages per minute, while the test function memory was set to 256MB, 512MB, and 1024MB. Through this feature, we are also able to regulate the maximum number of containers that can be launched to server incoming function requests. This number is the ratio between the node memory available to the Openwhisk process and the function memory used in each run. Hence, it needs to be viewed as the maximum number of available worker servers (or the c parameter in e.g. an M/M/c queuing model). The test function itself was the ehealth function presented in \cite{ehealth} and consists of a tensorflow-based AI model for inferring on a patient's state. The relevant information appears in Table 1. \par 
\noindent Load generation is performed through a relevant adapted client\cite{loadgenerator} that aims to adapt to the specifications of a FaaS system. These specifications include the need to get the result through asynchronous calls, meaning that the client can not block waiting for the request, but needs to poll afterwards for getting the result. This is the typical way through which Openwhisk serves the results of an executed function. Furthermore, the client is able to log different timestamps in the process, like initial client sample time, total response time, as well as process FaaS results in order to export provided statistics, such as wait time in the FaaS system, initialization time for the function container and pure function execution time (or service time).\par 

\noindent Other clients, like Apache Jmeter, which block while waiting for responses, can not follow the defined rate of requests unless the server is also able to sustain the desired throughput. However, a set rate load generator ensures a continuous and unrupted flow of requests. This approach allows for a stable and controlled load on the system under test, enabling to assert the desired message throughput.\par


\begin{table}[ht]
\caption{Testbed Data for Various Test Configurations}
\centering
\label{tab:table1}
\begin{adjustbox}{width=0.9\textwidth}
\begin{tabular}{|c|c|c|c|c|c|c|}
\hline
\multicolumn{4}{|c|}{Testbed} &  \multicolumn{3}{c|}{Test Function Memory (MB)} \\ \hline
 Testbed & Set Rate & Node Memory & Cores &  &  &  \\ 
  & (msg/min) & (GB) &  &256 & 512 & 1024\\ \hline 
\multirow{3}{*}{HUA} & \multirow{3}{*}{12,30,60} & \multirow{3}{*}{8} & \multirow{3}{*}{4} & \multicolumn{3}{c|}{Max Worker Containers}\\ \hline
\multirow{3}{*}{AWS} & \multirow{3}{*}{12,30,60} & \multirow{3}{*}{8} & \multirow{3}{*}{4} & 32 & 16 & 8\\\hline
\multirow{3}{*}{AZURE} & \multirow{3}{*}{12,30,60} & \multirow{3}{*}{1.5} & \multirow{3}{*}{8} & 32 & 16 & 8\\ \hline
 &  &  &  & 6 & 3 & 1\\ \hline
\end{tabular}
\end{adjustbox}

\end{table}

\noindent In order to eliminate any interference from initial cold starts on the results, a pre-warming run was conducted for each case. This step was necessary due to the inclusion of container initialization time in the function execution duration in Openwhisk, despite being reported separately. Once the warm containers were available following the pre-warm run, data collection took place during the main run for a duration of 600 seconds for each case. A total of 4890 samples were collected throughout the process and published by Kousiouris \cite{ehealth}.

\subsection{Raw Data Collection Presentation }

\noindent The outcomes of the experimental analysis are presented in this section, focusing primarily on the aspect of time from each experiment start. The influence of different variables (function memory, set rates, cluster type) on system performance across various scenarios, can be observed as well as its evolution as the experiment progresses. By examining these graphs, trends, patterns, and correlations can be identified, helping to understand how different variables affect system performance. \par

\subsubsection{HUA Testbed Results}

\noindent 
By examining the first series on the HUA testbed (Figure 2), it is evident that in the low rate case (i.e. 12 msg/min) the system is very stable, having a small wait time that is mainly functional (e.g. waiting for the incoming request to be forwarded across the various sub-components of the FaaS architecture). The amount of memory does not affect the function performance, since in all 3 memory variations the execution time is very similar (~6.5 seconds). If we apply Little's law in this case, with an incoming rate of 1 request per 5 seconds and a response time 6.5 seconds, we get an average of 1.3 customers in the system. This means that on average 2 max containers are needed for serving the functions in these scenarios and they run concurrently for about 30 percent of their time. A similar case applies for the medium rate, although in this case we have 1 request per 2 seconds and a slightly increased execution time of ~7 seconds leading to about 3.5 customers in the system and, thus, 4 max concurrent containers. 
However, in the case of the high rate (subfigures g-i), the system becomes unstable as is evident by the constantly rising wait time, although there are in principle 32,16 and 8 workers (max container slots) and this should be enough to accommodate the increased traffic (e.g. based on typical M/M/C formulas).  \par 

\begin{figure}
    \centering
    \begin{subfigure}[b]{0.32\textwidth}
        \includegraphics[width=\linewidth]{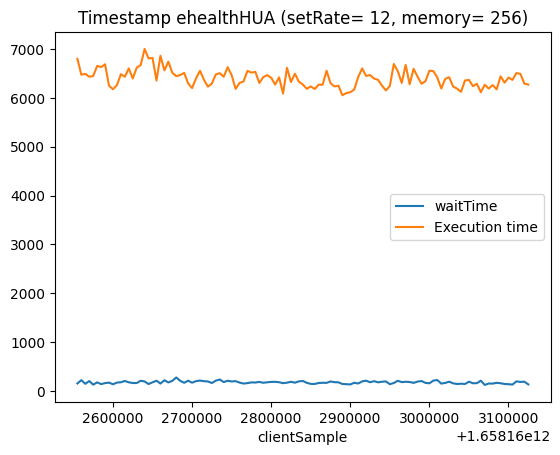}
        \caption{$setRate12\_mem256$}
        \label{fig:ehealthHUA_setRate 12-memory_256.png}
    \end{subfigure}
    \hfill
    \begin{subfigure}[b]{0.32\textwidth}
        \includegraphics[width=\linewidth]{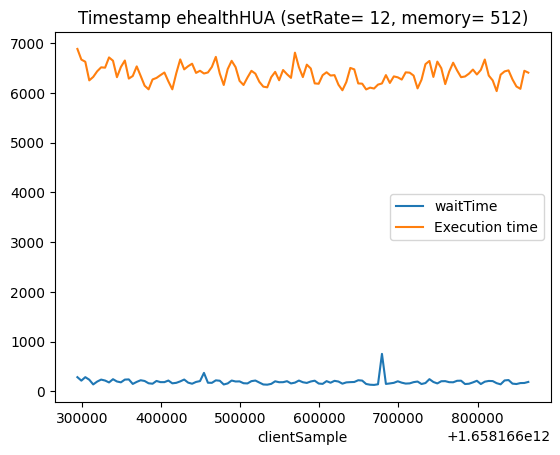}
        \caption{$setRate12\_mem512$}
        \label{fig:ehealthHUA_setRate 12-memory 512}
    \end{subfigure}
    \hfill
    \begin{subfigure}[b]{0.32\textwidth}
        \includegraphics[width=\linewidth]{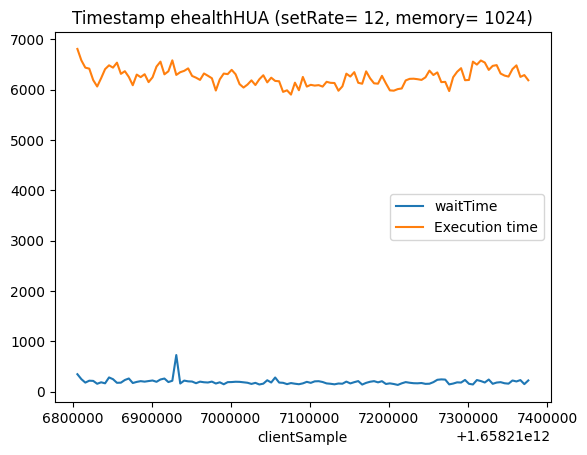}
        \caption{$setRate12\_mem1024$}
        \label{fig:ehealthHUA_setRate 12-mem1024}
    \end{subfigure}
    \\
    \begin{subfigure}[b]{0.32\textwidth}
        \includegraphics[width=\linewidth]{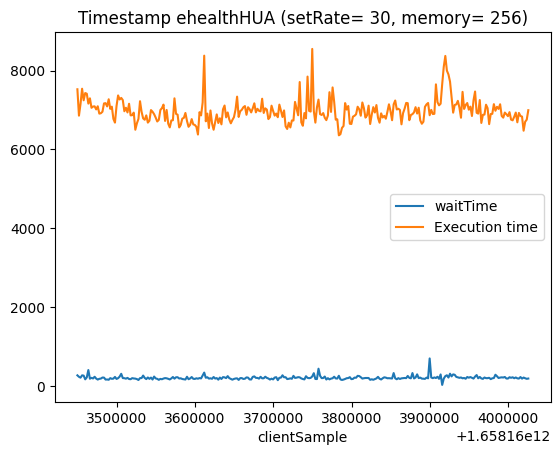}
        \caption{$setRate30\_mem256$}
        \label{fig:ehealthHUA_setRate 30-memory 256}
    \end{subfigure}
    \hfill
    \begin{subfigure}[b]{0.32\textwidth}
        \includegraphics[width=\linewidth]{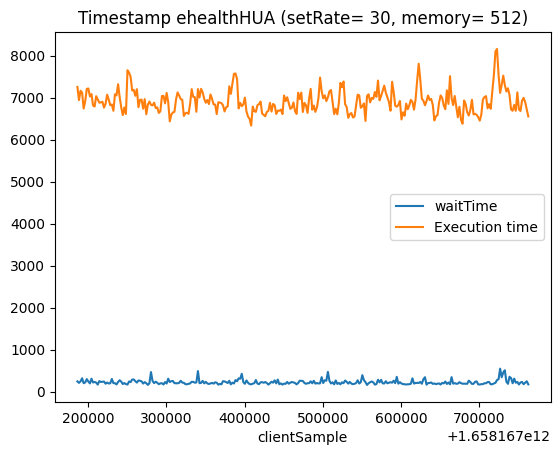}
        \caption{$setRate30\_mem512$}
        \label{fig:ehealthHUA_setRate 30-memory 512}
    \end{subfigure}
    \hfill
    \begin{subfigure}[b]{0.32\textwidth}
        \includegraphics[width=\linewidth]{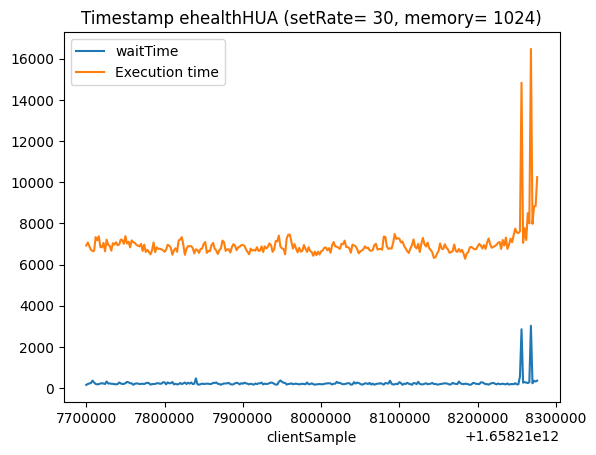}
        \caption{$setRate30\_mem1024$}
        \label{fig:ehealthHUA_setRate 30-memory 1024}
    \end{subfigure}
    \\
    \begin{subfigure}[b]{0.32\textwidth}
        \includegraphics[width=\linewidth]{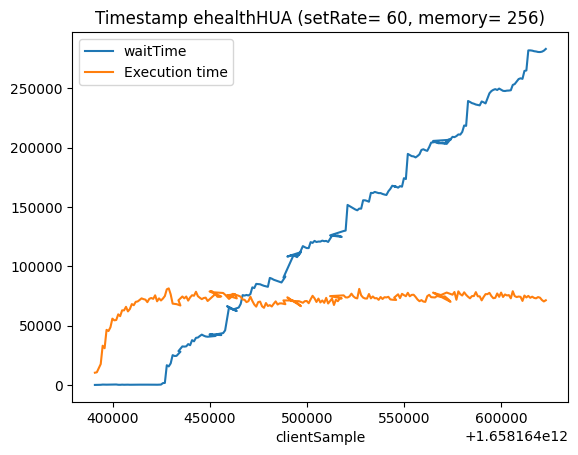}
        \caption{$setRate60-mem256$}
        \label{fig:ehealthHUA_setRate 60-memory 256}
    \end{subfigure}
    \hfill
    \begin{subfigure}[b]{0.32\textwidth}
        \includegraphics[width=\linewidth]{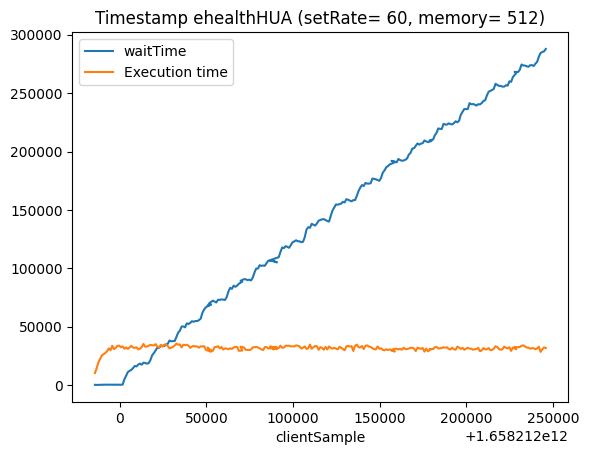}
        \caption{$setRate60\_mem512$}
        \label{fig:ehealthHUA_setRate 60-memory 512}
    \end{subfigure}
    \hfill
    \begin{subfigure}[b]{0.32\textwidth}
        \includegraphics[width=\linewidth]{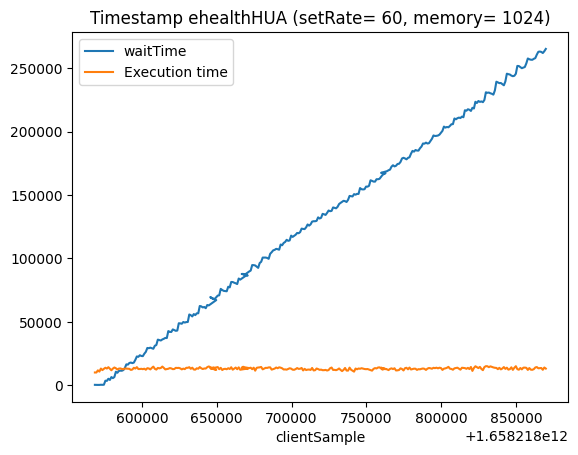}
        \caption{$setRate60\_mem1024$}
        \label{fig:ehealthHUA_setRate 60-memory_1024}
    \end{subfigure}
    \caption{Performance plots of ehealthHUA}
    \label{fig:ehealthHUA}
\end{figure}

\noindent However there is a significant difference in the case of FaaS compared to the application of queuing models in other domains (e.g. a supermarket or a calling center model). As seen in Fig. 1, a FaaS worker node internally divides its computational capacity into smaller slots (or container slots). Thus, a number of function containers can be squeezed in the same worker node depending on memory availability. This means that the overall resources of the node are now shared by the concurrent processes which leads to significant degradation in the function execution performance, due to direct (cpu time) or indirect (cache contamination) resource sharing. \par

\noindent In most cases queuing model approaches assume that the service rate of the system is not dependent on the system state (in this case the number of concurrently served customers/containers in the system) but on a distribution that relates primarily to differences in the function execution itself. In a supermarket analogy, one customer may require more time to be served compared to another (because they have a larger cart) but the time it takes to be served does not depend on the number of tellers in the supermarket. In reality, what happens is that the serving time (function execution duration) will also depend on the amount of tellers in the system (because they waste time chatting to each other for example). \par
\noindent For quantifying this increase, one can observe Fig. 2 (g-i), in which the system is operating using the max available containers. In the case of Fig. 2g the function execution duration increased from 6.5 seconds to ~ 70 seconds. In the case of Fig. 2(h), we have 16 max containers (and a service time of ~35 seconds) while in Fig. 2(i) we have 8 max containers and a service time of ~12 seconds. So it seems to be following a rather linear reduction based on the percentage of cpu core time assigned to each container, once the container numbers exceed the amount of available cores (4 in the case of the HUA testbed). This is reasonable since after passing the ratio of 1 container per core the core time is split between the competing containers. \par
\noindent This leads us to a chicken and egg problem when trying to apply queueing models in such environments. How can we estimate how many customers in the system exist at any given time, given that this depends on the service rate, and on the other hand, how can we estimate the service rate when it depends on the customers in the system. An iterative approach could be potentially followed, based on split time windows in order also to adapt to incoming traffic variations, in which either the execution duration or the customers in the system are monitored and from this parameter we can estimate the remaining ones. \par

\subsubsection{AWS Testbed Results}
\noindent For the AWS case (Fig. 3a-f), we can observe similar behavior, with two exceptions. The 256 MB case consistently fails due to function memory being too low. The reason, compared to the HUA testbed, is that in this case we are using Kubernetes container management system, which kills the container even if a slight memory violation occurs. On the HUA testbed, we use the Docker Engine, which is not that strict. \par

\subsubsection{Probability of Failures Due to Image Pulling Bottlenecks}
  
\noindent Furthermore, we can observe a number of anomalies in the AWS high rates (Fig 3f-g). While the system appears to be stable (in the sense that it has an increased but manageable service time), there are a number of spikes in waiting time, as well as an increased number of failures in the requests. In order to further investigate this phenomenon, we broke down the responses to successful and unsuccessful ones, appearing in Fig. 4. After examination, it was determined that these failures were due to numerous concurrent requests for pulling and launching the function container image and can be attributed to the image registry and the according storage system. A relevant ErrImagePull event was generated due to this spike in image pulling, which was caused by the spike in cold starts. Cold starts are treated in literature significantly, however what this analysis shows is that they also need to be linked not only with extended initialization times, but also with higher probabilities of failure. Thus, it should be added as a stage in a kind of Markov chain analysis. \par  

\begin{figure}[htbp]
    \centering
    \begin{subfigure}[b]{0.32\textwidth}
        \includegraphics[width=\linewidth]{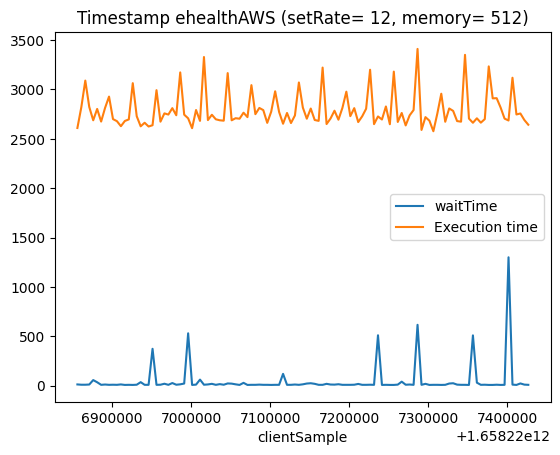}
        \caption{$setRate12\_mem512$}
        \label{fig:ehealthAWS_setRate12-memory 512}
    \end{subfigure}
    \hfill
    \begin{subfigure}[b]{0.32\textwidth}
        \includegraphics[width=\linewidth]{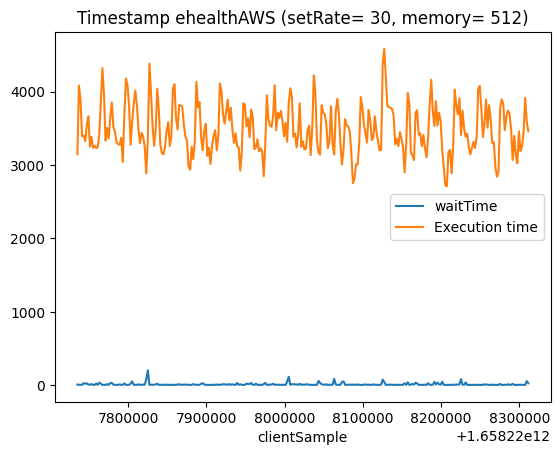}
        \caption{$setRate30\_mem512$}
        \label{fig:ehealthAWS_setRate30_mem512}
    \end{subfigure}
    \hfill
    \begin{subfigure}[b]{0.32\textwidth}
        \includegraphics[width=\linewidth]{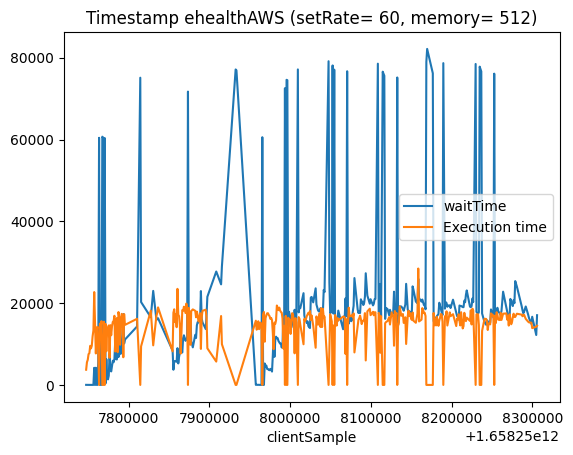}
        \caption{$setRate60\_mem512$}
        \label{fig:ehealthAWS_setRate60-memory 512}
    \end{subfigure}
    \\
    \begin{subfigure}[b]{0.32\textwidth}
        \includegraphics[width=\linewidth]{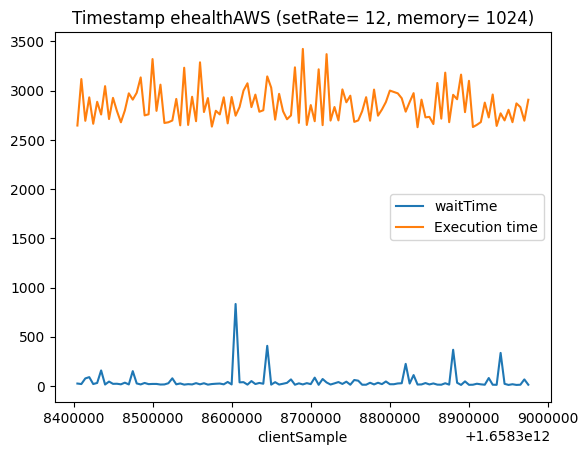}
        \caption{$setRate12\_mem1024$}
        \label{fig:ehealthAWS_setRate12-memory 1024}
    \end{subfigure}
    \hfill     
    \begin{subfigure}[b]{0.32\textwidth}
        \includegraphics[width=\linewidth]{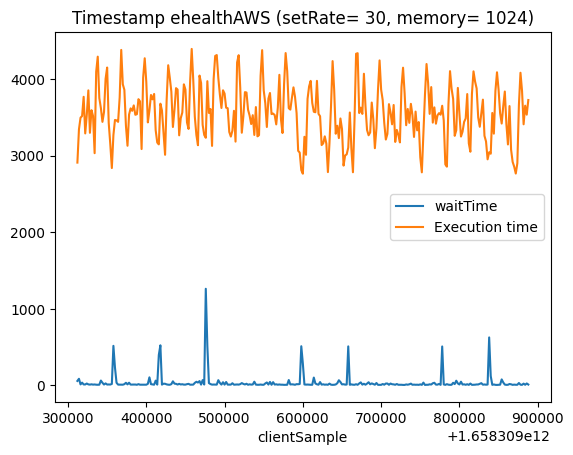}
        \caption{$setRate30\_mem1024$}
        \label{fig:ehealthAWS_setRate30-memory 1024}
    \end{subfigure}
    \hfill
    \begin{subfigure}[b]{0.32\textwidth}
        \includegraphics[width=\linewidth]{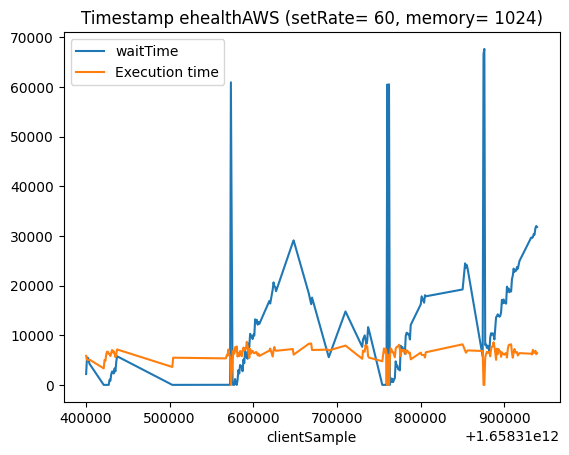}
        \caption{$setRate60\_mem1024$}
        \label{fig:}
    \end{subfigure}
    \caption{Performance plots of ehealthAWS}
    \label{fig:ehealthAWS}
\end{figure}

\begin{figure} [htbp]
  \centering
  \begin{subfigure}[b]{0.45\textwidth}
    \includegraphics[width=\textwidth]{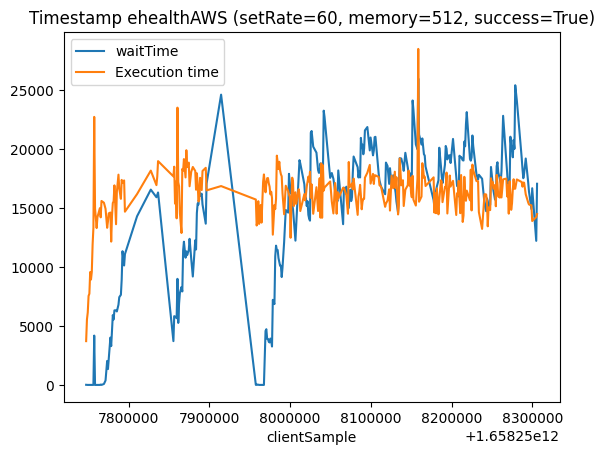}
    \caption{success true}
  \end{subfigure}   
  \hfill
  \begin{subfigure}[b]{0.45\textwidth}
    \includegraphics[width=\textwidth]{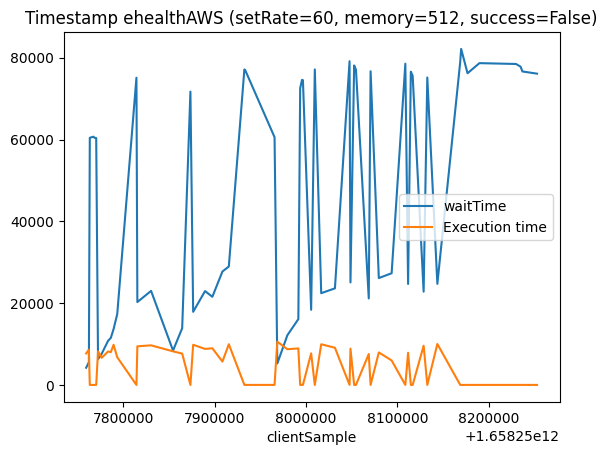}
    \caption{success False}
  \end{subfigure}
  \caption{Behavior of the ehealthAWS with setRate=60 and memory=512}
  \label{fig:Figure4}
\end{figure}


\subsubsection{Azure Testbed Results}
\begin{figure} [htbp]
    \centering
    \begin{subfigure}[b]{0.32\textwidth}
        \includegraphics[width=\linewidth]{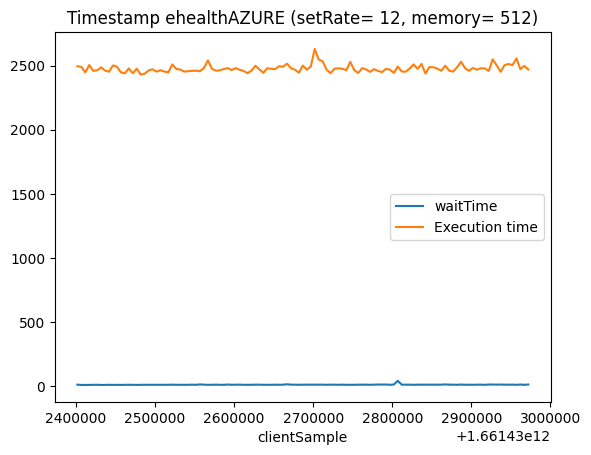}
        \caption{$setRate12\_mem512$}
        \label{fig:ehealthAZURE_setRate12-memory 512}
    \end{subfigure}
    \hfill
    \begin{subfigure}[b]{0.32\textwidth}
        \includegraphics[width=\linewidth]{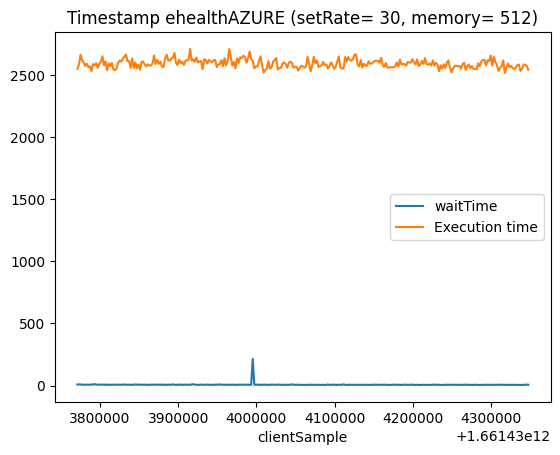}
        \caption{$setRate30\_mem512$}
        \label{fig:ehealthAZURE_setRate30_mem512}
    \end{subfigure}
    \hfill
    \begin{subfigure}[b]{0.32\textwidth}
        \includegraphics[width=\linewidth]{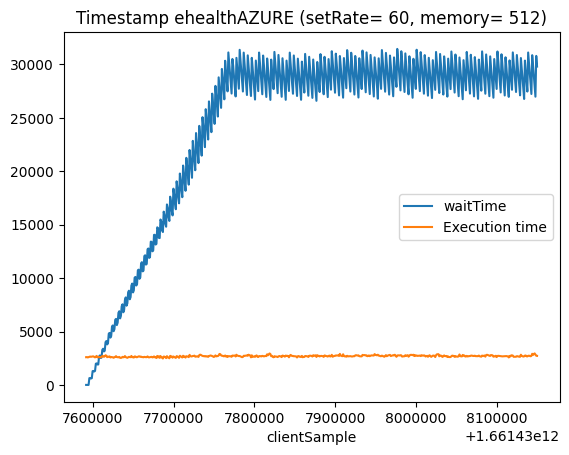}
        \caption{$setRate60\_mem512$}
        \label{fig:ehealthAZURE_setRate60-memory 512}
    \end{subfigure}
    \\
    \begin{subfigure}[b]{0.32\textwidth}
        \includegraphics[width=\linewidth]{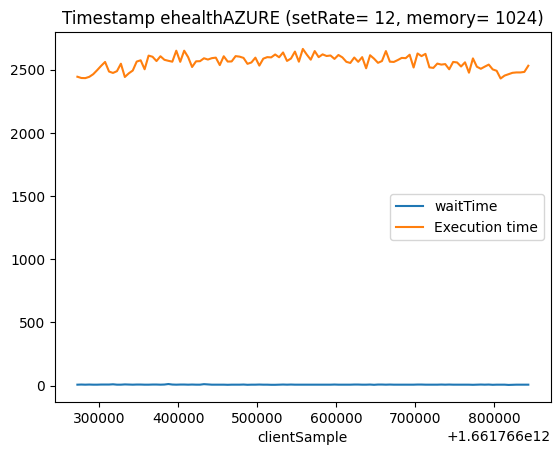}
        \caption{$setRate12\_mem1024$}
        \label{fig:ehealthAZURE_setRate12-memory 1024}
    \end{subfigure}
    \hfill     
    \begin{subfigure}[b]{0.32\textwidth}
        \includegraphics[width=\linewidth]{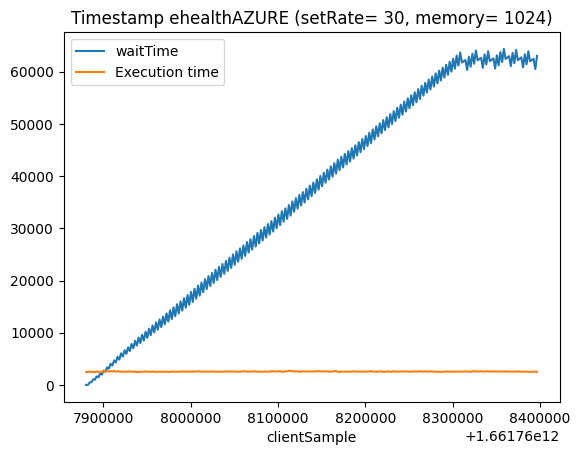}
        \caption{$setRate30\_mem1024$}
        \label{fig:ehealthAZURE_setRate30-memory 1024}
    \end{subfigure}
    \hfill
    \begin{subfigure}[b]{0.32\textwidth}
        \includegraphics[width=\linewidth]{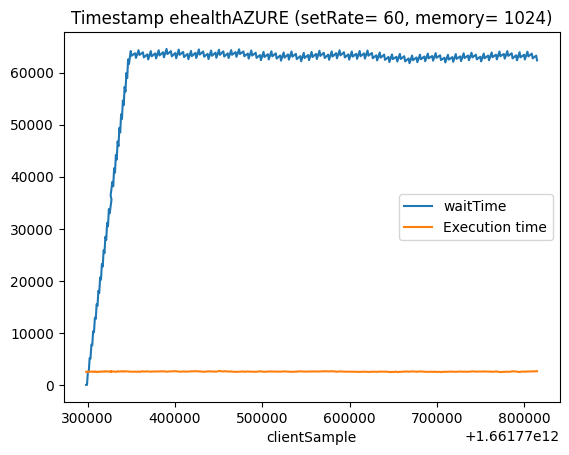}
        \caption{$setRate60\_mem1024$}
        \label{fig:ehealthAZURE_setRate60-memory 1024}
    \end{subfigure}
    \caption{Performance plots of ehealthAZURE}
    \label{fig:ehealthAZURE}
\end{figure}

\noindent Respectively, in the Azure testbed, the results appear in Fig. \ref{fig:ehealthAZURE}. In the low rate case, the system is stable and performing slightly better than AWS, given that the VM is based on a slightly more advanced cpu. Here a number of comparisons can be performed. Comparing between the rate 30-512 MB cases, it can be noted that the AWS execution time is raised to ~ 3.5 seconds. On the other hand, the Azure case is more stable around 2.5 seconds (as the baseline value in the low rate scenario).  In 30-1024 MB case, AWS still maintains the ~3.5 seconds execution and Azure the 2.5 one, however due to the fact that Azure has the restriction of only 1 worker, requests are queued given that they arrive at a rate of 1 every 2 seconds. The key take away from these testbed results is that, if we want to maintain a very stable execution time, then we need to do it at the expense of higher waiting time. There is also one other peculiarity in the graphs that will be explained in detail in Section 4.1. \par

\section{Results, Analysis and Discussion}
\label{Results}

\noindent In this section, we further analyse the results depicted in the provided graphs of section 3, as well as we provide some insights into the back-end processes. More importantly, we highlight the differences between the results of the cases and discuss them in an attempt to also explain them and, thus, support future researchers and practitioners that will explore FaaS systems in cloud/edge computing environments. \par

\subsection{Cut-off points and Load Generation Analysis}\par

\noindent From the AWS and Azure cases, some peculiarities were observed in the exported graphs. For example, in the case of Azure (Fig. \ref{fig:ehealthAZURE} c-e-f, the queue seems to be constantly rising, until a point in which it becomes stable. Given that the arrival rate is higher than the departure rate from all servers, this queuing time should be constantly rising. Initially it was considered that it might be a problem of the load generator.  FaaS platforms typically do not allow blocking calls for a delay larger than a specific limit, thus async invocations should be performed and the result acquired through a polling process.\par
\noindent Thus, a high number of actively monitored functions could put a large strain on the load generator, not enabling it to meet the desired rate. Furthermore, the load generator is built in such a way that it logs only the requests from which a result was obtained from the FaaS platform.\par
\begin{figure}[htbp]
  \centering
  
  \begin{subfigure}[b]{0.3\textwidth}
    \centering
    \includegraphics[width=\textwidth]{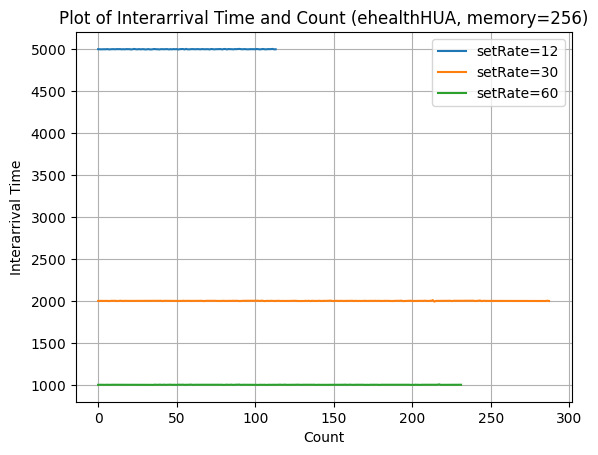}
    \caption{ehealthHUA\_Memory256}
    \label{fig:ehealthHUA_Memory 256}
  \end{subfigure}
  \hfill
  \begin{subfigure}[b]{0.3\textwidth}
    \centering
    \includegraphics[width=\textwidth]{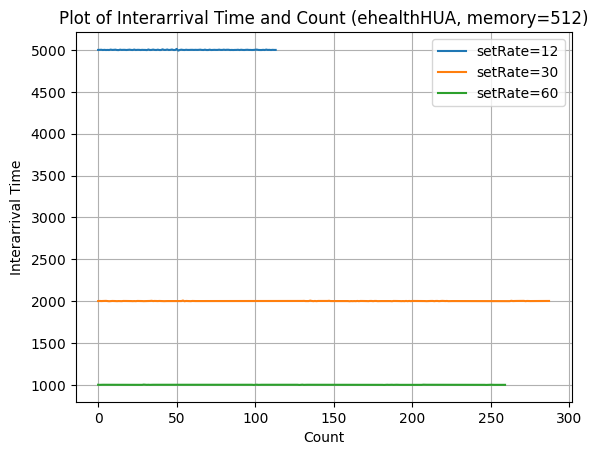}
    \caption{ehealthHUA\_Memory512}
    \label{fig:ehealthHUA_Memory 512}
  \end{subfigure}
  \hfill
  \begin{subfigure}[b]{0.3\textwidth}
    \centering
    \includegraphics[width=\textwidth]{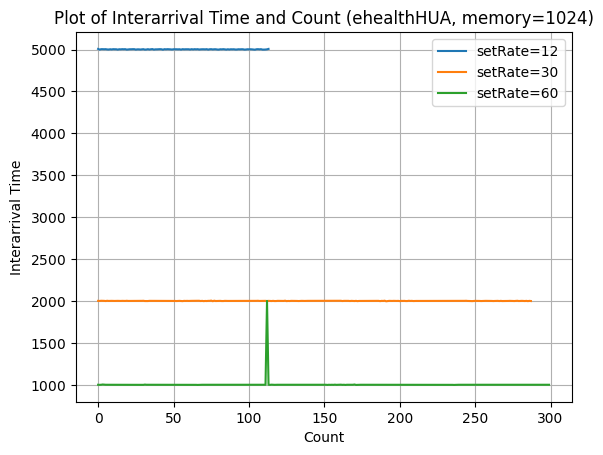}
    \caption{ehealthHUA\_Memory1024}
    \label{fig:ehealthHUA_Memory 1024}
  \end{subfigure}
  
  \caption{Inter-arrival time (seconds) between processed request samples in ehealthHUA}
  \label{fig:Figure6}
\end{figure}

\begin{figure}[htbp]
    \centering
    
    \begin{subfigure}[b]{0.45\textwidth}
        \centering
        \includegraphics[width=\textwidth]{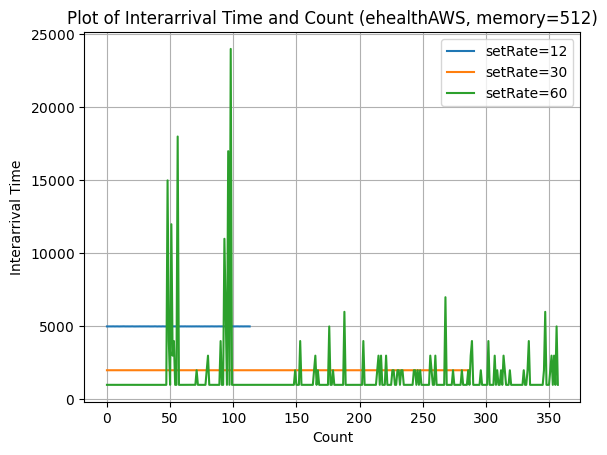}
        \caption{ehealthAWS\_Memory 512}
    \end{subfigure}
    \hfill
    \begin{subfigure}[b]{0.45\textwidth}
        \centering
        \includegraphics[width=\textwidth]{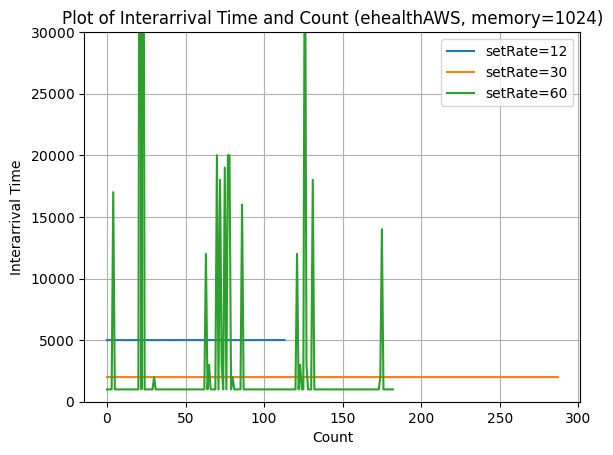}
        \caption{ehealthAWS\_Memory 1024}
    \end{subfigure}
    
    \caption{Inter-arrival time (seconds) between processed request samples in ehealthAWS}
    \label{fig:Figure7}
\end{figure}

\begin{figure}[htbp]
    \centering    
    \begin{subfigure}[b]{0.45\textwidth}
        \centering
        \includegraphics[width=\textwidth]{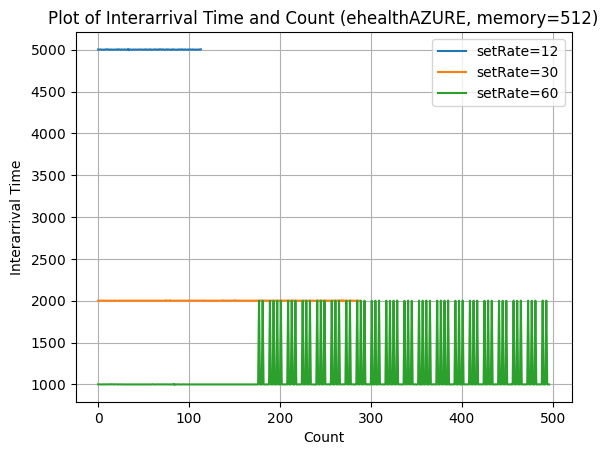}
        \caption{ehealthAZURE\_Memory 512}
    \end{subfigure}
    \hfill
    \begin{subfigure}[b]{0.45\textwidth}
        \centering
        \includegraphics[width=\textwidth]{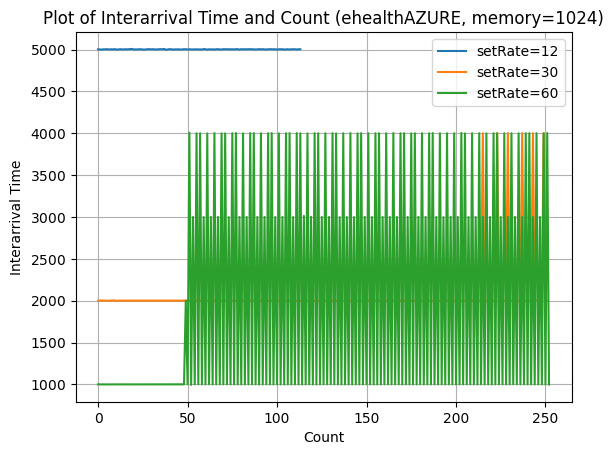}
        \caption{ehealthAZURE\_Memory 1024}
    \end{subfigure}
    \caption{Inter-arrival time (seconds) between processed request samples in ehealthAzure}
    \label{fig:Figure8}
\end{figure}

 
\noindent In order to evaluate the performance of the load generator, each request timestamp was logged at the client side and compared to the target rate. No deviation was found on that side, following the desired request rate. Then, we plotted the received requests in the three scenarios, that appear in Figs.  \ref{fig:Figure7} and \ref{fig:Figure8}. The samples are sorted based on the timestamp of the initial request (x axis) and the time difference between sample n and (n-1) (y axis). Hence, a consistent request/response rate should be observed, similar to the cases of Figure \ref{fig:Figure6}, thus a straight line indicating the set rate in requests per minute. The aforementioned case in the HUA testbed appears to be working as expected. The fact that we have less samples in the HUA case in the 60 rate scenario can be explained due to the high execution time in this case. The load generator waits for the 600 seconds of each experiment duration and then proceeds to calculate the samples arrived up to then. Thus, samples that arrive after that mark are neglected, hence the low sample count collection in the high rate/high delay case.\par
\noindent However, in the cases of Amazon and especially Azure, the samples in the high rate case (and in some cases in the medium rate towards the end) do not follow the specified rate. Furthermore, this difference appears to be very specific, i.e. instead of having samples every 1 second we get a specific sequence like 1,4,1,3,1 etc. This was due to the fact that in FaaS platforms there are two cut-off points that can be set. One cut-off point is at the overall level, indicating that the platform should reject invocations higher than e.g. 60 per minute. Another cut-off point is the fact that one can determine the maximum number of concurrent, active  invocations. It is due to this limit that this queue levelling should be attributed. Once this limit is reached, the platform rejects the next request until one of the previous ones has been finished and that is why the interarrival times of processed requests are always at a multiple of 1. In the case of HUA this limit was set much higher (400 invocations per minute) than the AWS and Azure cases in which the specific cut-off point was set to 60 invocations per minute, hence the difference in the processed request sample plots. \par


\noindent These details were gathered and calculated in Tables \ref{tab:huarates}, \ref{tab:awsrates} and \ref{tab:azurerates} for the three cases respectively. What is mostly interesting in this case is to calculate from the total experiment duration which were the acquired samples compared to the formally foreseen ones. This can help us determine the effective rate of requests, i.e. the actual part of the arrival rate that indeed enters the system. Furthermore, the probability of a request for a given scenario to enter the system is important. For example, in cases where we need to foresee such a state (such as in Markov chains), as well as the probability of transitioning to that state.
The difference for higher rejection in the case of AWS compared to Azure may be attributed  to the fact that AWS due to the concurrency overheads and the higher number of allowed containers has a higher execution time, thus the rate with which requests leave the system is lower. This leads to more requests being rejected due to the waiting queue being full. 
\par

\begin{table}[htbp]
 \caption{Effective Rate and Request Acceptance Probability for test\_name = ehealthHUA }
  \centering
  \begin{adjustbox}{width=0.9\textwidth}
  \begin{tabular}{|c|c|c|c|c|c|}
    \hline
    \textbf{Target Rate (request/sec)} & \textbf{Memory (MB)} & \textbf{Actual Samples} &  \textbf{Expected Samples} & \textbf{Effective rate} & \textbf{Probability of }\\ 
     & &  &   &  & \textbf{accepted requests}\\ \hline
    0.2 (12 request/min) & 256 & 114 & 114 & 0.199 & 1 \\ \hline 
    0.2 (12 request/min) & 512 & 114 & 114 & 0.199 & 1 \\ \hline  
    0.2 (12 request/min) & 1024 & 114 & 114 & 0.199 & 1 \\ \hline  
    \hline
    0.5 (30 request/min) & 256 & 288 & 288 & 0.499 & 1 \\ \hline  
    0.5 (30 request/min) & 512 & 288 & 288 &  0.499 & 1 \\ \hline  
    0.5 (30 request/min) & 1024 & 288 & 288 & 0.499 & 1 \\ \hline  
    \hline
    1 (60 request/min) & 256 & 232 & 232 & 0.999 & 1 \\ \hline 
    1 (60 request/min) & 512 & 260 & 260 & 0.999 & 1 \\ \hline 
    1 (60 request/min) & 1024 & 300 & 301 & 0.995 & 0.995\\ \hline  
  \end{tabular}
  \end{adjustbox}
  \label{tab:huarates}
\end{table}

\begin{table}[htbp]
   \caption{Effective Rate and Request Acceptance Probability for test\_name = ehealthAWS}
  \centering
  \begin{adjustbox}{width=0.9\textwidth}
  \begin{tabular}{|c|c|c|c|c|c|}
    \hline
    \textbf{Target Rate (request/sec)} & \textbf{Memory (MB)} & \textbf{Actual Samples} &  \textbf{Expected Samples} & \textbf{Effective rate} & \textbf{Probability of }\\ 
     & &  &   &  & \textbf{accepted requests}\\ \hline
    0.2 (12 request/min) & 512 & 114 & 114 & 0.199 & 0.999\\ \hline
    0.2 (12 request/min) & 1024 & 114 & 114 & 0.199 & 1\\ \hline
    \hline
    0.5 (30 request/min) & 512 & 288 & 288 & 0.499 & 1\\ \hline 
    0.5 (30 request/min) & 1024 & 288 & 288 & 0.499 & 1\\ \hline 
    \hline
    1 (60 request/min) & 512 & 358 & 558 & 0.641 & 0.641\\ \hline 
    1 (60 request/min) & 1024 & 183 & 539 & 0.339 & 0.339\\ \hline 
  \end{tabular}
  \end{adjustbox} 
    \label{tab:awsrates}
\end{table}

\begin{table}[htbp]
  \caption{Effective Rate and Request Acceptance Probability for test\_name = ehealthAZURE}
  \centering 
  \begin{adjustbox}{width=0.9\textwidth}
  \begin{tabular}{|c|c|c|c|c|c|}
    \hline
    \textbf{Target Rate (request/sec)} & \textbf{Memory (MB)} & \textbf{Actual Samples} &  \textbf{Expected Samples} & \textbf{Effective rate} & \textbf{Probability of }\\ 
     & &  &   &  & \textbf{accepted requests}\\ \hline
    0.2 (12 request/min) & 512 & 114 & 114 & 0.199 & 1 \\ \hline
    0.2 (12 request/min) & 1024 & 114 & 114 & 0.199 & 1 \\ \hline
    \hline
    0.5 (30 request/min) & 512 & 288 & 288 & 0.499 & 1 \\ \hline  
    0.5 (30 request/min) & 1024 & 252 & 288 & 0.499 & 0.976\\ \hline  
    \hline
    1 (60 request/min) & 512 & 497 & 558 & 0.8902 & 0.8902\\ \hline  
    1 (60 request/min) & 1024 & 253 & 516 & 0.4901 & 0.4901\\ \hline  
  \end{tabular}
  \end{adjustbox}
  \label{tab:azurerates}
\end{table}

\subsection{Cost Model Consideration for Trade-off between Waiting and Execution Time}

\subsubsection{Waiting and Execution Time Trade-offs}

\noindent The distribution of function execution time data for each testbed, as presented in Figures 9-11, provides valuable insights into the performance characteristics and potential container overheads or resource sharing penalties. This is influenced by various factors, including the management and execution of containers, as well as the system configuration and design choices.\par
\noindent This analysis allows us to observe the stability of the executions, as well as use them as potential future inputs into identifying the distribution types to be used in modelling approaches. It can also give insights into the split of the total response time between waiting and execution, based on the cluster setup. As an example, if we investigate the graphs of the rate 60-memory 512 case in the AWS testbed (Fig.10(b)), we can observe that it portrays a rather increased execution time, as well as deviation, given that the distribution spans across a large interval (2500-20000 milliseconds). This is reasonable since in this case we can expect that a significantly high number of containers is concurrently active. On the other hand, the Azure rate 60-memory 512 case (max 3 concurrent containers) (Fig.10(c)) is very much concentrated around a small interval (2500-2900 milliseconds). On the waiting time aspect, things are quite the opposite, with AWS portraying the majority of delays up to 20000 milliseconds while Azure over 20000 milliseconds. This is reasonable since with fewer container slots, requests need to be queued to find one available. 


\begin{figure} [b]
    \centering
    \includegraphics[width=0.3\textwidth]{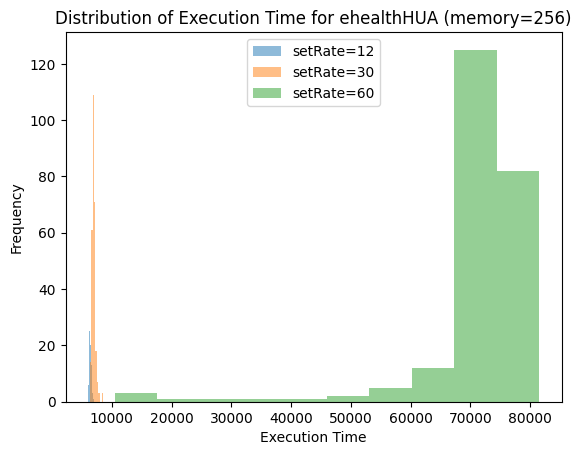}
    \caption{Distribution of Execution Time for Memory 256 }
    \label{fig:Figure9}
\end{figure}

\begin{figure}
  \centering 
  \begin{subfigure}[b]{0.3\textwidth}
    \centering
    \includegraphics[width=\textwidth]{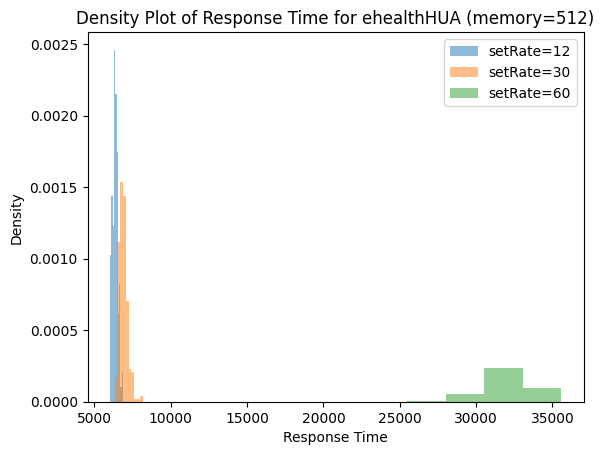}
    \caption{ehealthHUA}
  \end{subfigure}
  \hfill
  \begin{subfigure}[b]{0.3\textwidth}
    \centering
    \includegraphics[width=\textwidth]{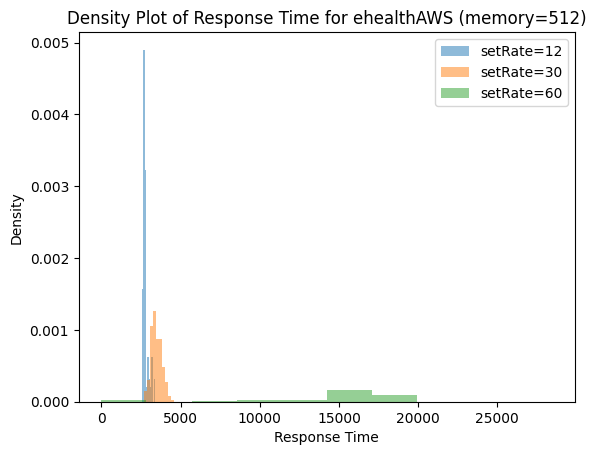}
    \caption{ehealthAWS}
  \end{subfigure}
  \hfill
  \begin{subfigure}[b]{0.3\textwidth}
    \centering
    \includegraphics[width=\textwidth]{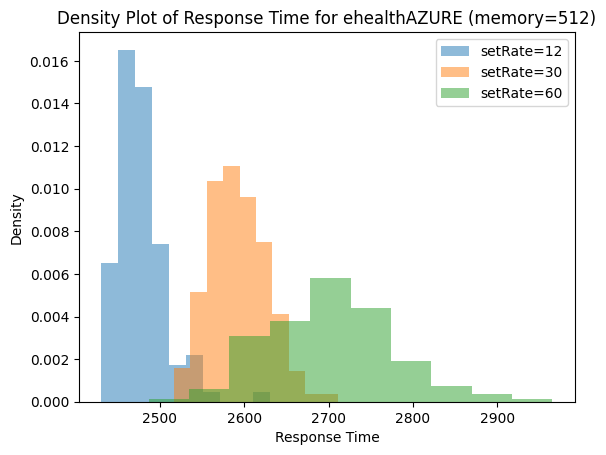}
    \caption{ehealthAZURE}
  \end{subfigure}  
  \caption{Distribution of Execution Time for Memory 512}
  \label{fig:Figure10}
\end{figure}

\begin{figure}
  \centering  
  \begin{subfigure}[b]{0.3\textwidth}
    \centering
    \includegraphics[width=\textwidth]{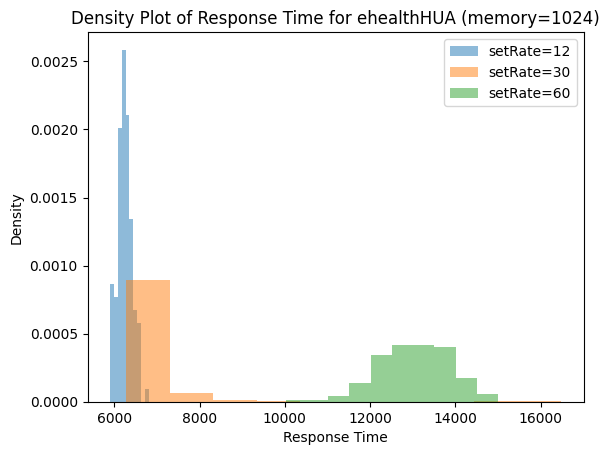}
       \caption{ehealthHUA}
  \end{subfigure}
  \hfill
  \begin{subfigure}[b]{0.3\textwidth}
    \centering
    \includegraphics[width=\textwidth]{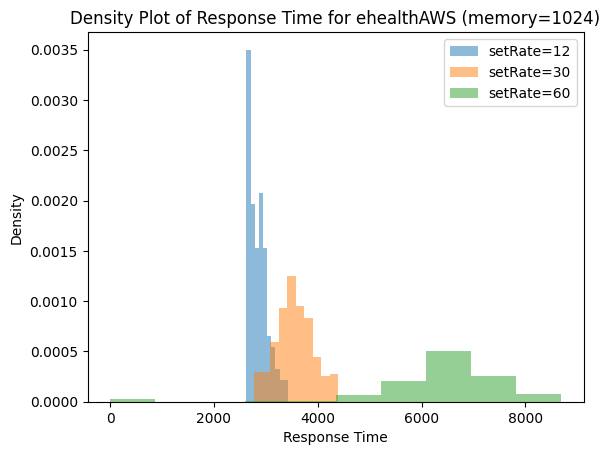}
   \caption{ehealthAWS}
  \end{subfigure}
  \hfill
  \begin{subfigure}[b]{0.3\textwidth}
    \centering
    \includegraphics[width=\textwidth]{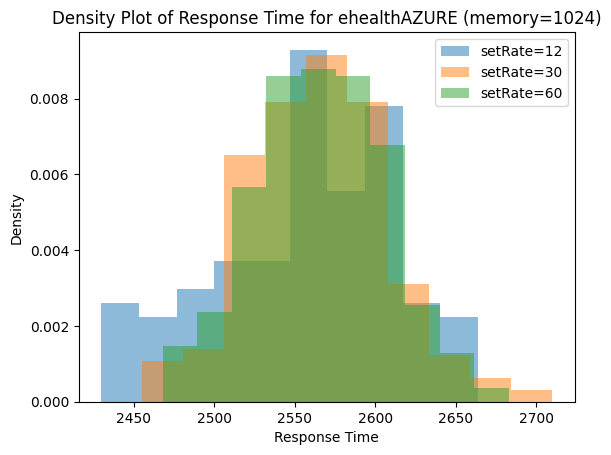}
     \caption{ehealthAZURE}
  \end{subfigure}
  
  \caption{Distribution of Execution Time for Memory 1024 }
  \label{fig:Figure11}
\end{figure}

\begin{figure}
    \centering
    \includegraphics[width=0.3\textwidth]{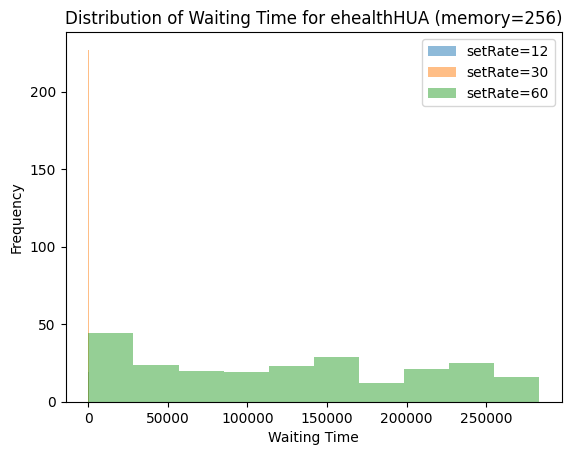}
    \caption{Distribution of Waiting Time for Memory 256 }
    \label{fig:Figure12}
\end{figure}

\begin{figure}
  \centering 
  \begin{subfigure}[b]{0.3\textwidth}
    \centering
    \includegraphics[width=\textwidth]{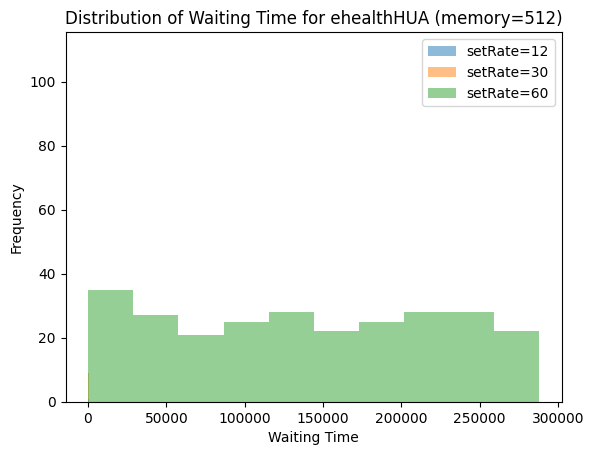}
    \caption{ehealthHUA}
  \end{subfigure}
  \hfill
  \begin{subfigure}[b]{0.3\textwidth}
    \centering
    \includegraphics[width=\textwidth]{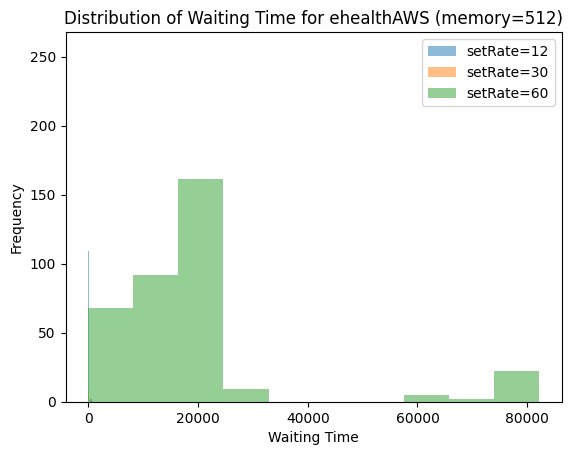}
    \caption{ehealthAWS}
  \end{subfigure}
  \hfill
  \begin{subfigure}[b]{0.3\textwidth}
    \centering
    \includegraphics[width=\textwidth]{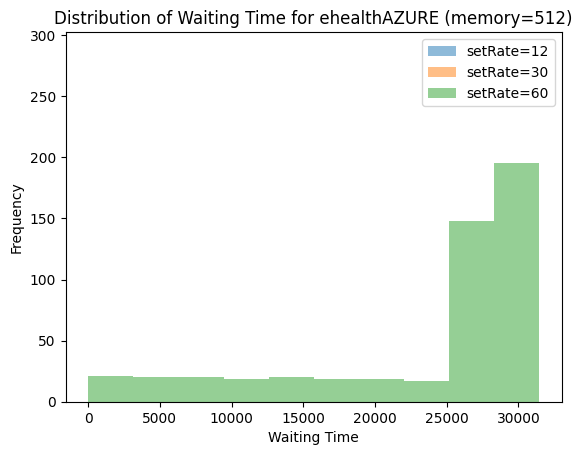}
    \caption{ehealthAZURE}
  \end{subfigure}
  
  \caption{Distribution of Waiting Time for Memory 512 }
  \label{fig:Figure13}
\end{figure}

\begin{figure}
  \centering  
  \begin{subfigure}[b]{0.3\textwidth}
    \centering
    \includegraphics[width=\textwidth]{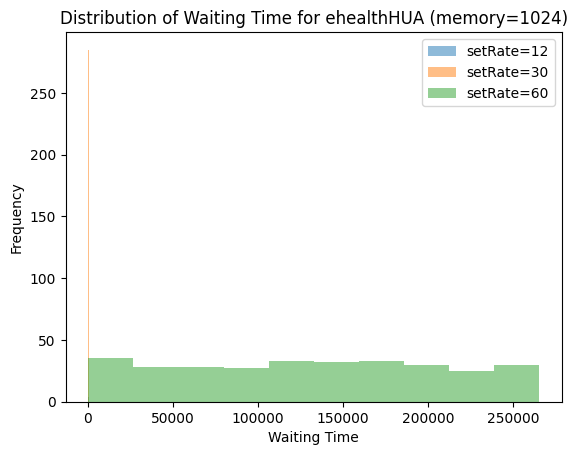}
    \caption{ehealthHUA}
  \end{subfigure}
  \hfill
  \begin{subfigure}[b]{0.3\textwidth}
    \centering
    \includegraphics[width=\textwidth]{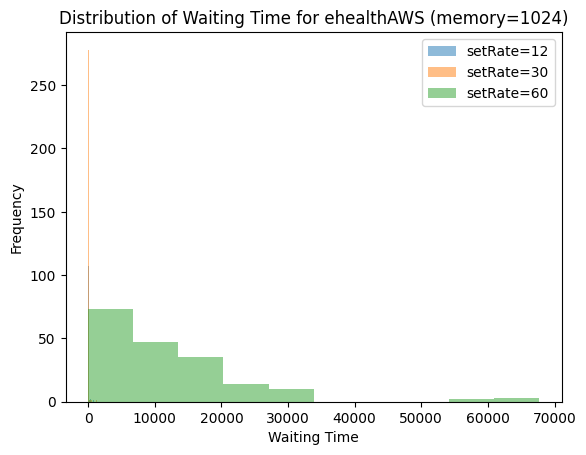}
    \caption{ehealthAWS}
  \end{subfigure}
  \hfill
  \begin{subfigure}[b]{0.3\textwidth}
    \centering
    \includegraphics[width=\textwidth]{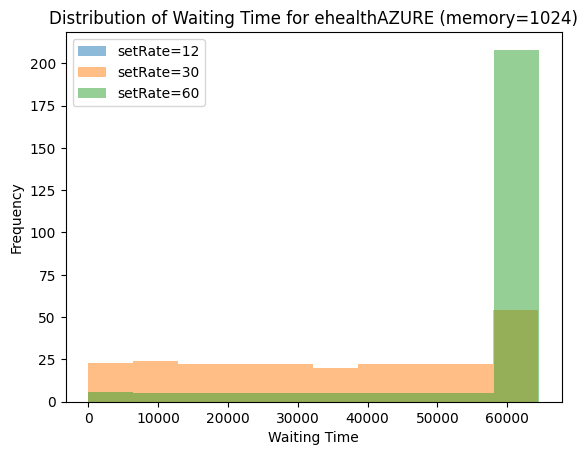}
    \caption{ehealthAZURE}
  \end{subfigure}
  
  \caption{Distribution of Waiting Time for Memory 1024}
  \label{fig:Figure14}
\end{figure}

\subsubsection{Cost Model Consideration for Waiting Time vs Execution Time}

\noindent The cost model for Function as a Service (FaaS) is typically determined based on the execution time of a given function and the amount of memory it allocates for its container\cite{amazon}. However, in the context of overloaded clusters, this model may be violated. Under increased load, customers may end up paying more for a degraded service quality due to the concurrency overheads. To prevent such violations, it is necessary to configure the system in a way that prioritizes maintaining a stable execution time, even at the expense of higher waiting time, as seen in the previous section. This ensures that customers are not penalized both with higher costs and a suboptimal user experience.\par

\noindent In order to compare between the scenarios and to check how different configurations and loads affect the experienced QoS, Table 5 was created. In this, we consider as baseline the HUA low rate scenario. Then, we calculate the percent differences ((T-Tbaseline)/Tbaseline) of each according time of the other scenarios. Hence, a number of insights can be extracted that can aid in two directions. On one hand, to compare the different available clusters in case of routing requests between them. On the other hand, they can be used to compare configurations and baseline abilities of each cluster. 
For example, the main concurrency problems start from a ratio of containers / cores $>$ 1. However, in the case of the AWS 60 rate for 8 max containers (double the number of the available cores on the node), although we get almost double the execution time compared to the AWS low rate case, the achieved execution time is almost identical to the baseline HUA scenario that is used as the reference.  This gives us an overall good trade-off between execution and wait time, indicated by the fact that this setup has the lowest degradation in the total response time (174\%) of the high rate case.
On the other hand, if we want to keep a very consistent execution time, then configurations, such as the Azure testbed, can be used, In these setups, the ratio of 1 container per core is not violated, thus indicating a stable improvement of around 60\% compared to the baseline HUA execution. However, the user should also be willing to trade this stability with a higher wait and total time. \par

\begin{table}[ht]
\caption {Comparison of Testbed Data for Various Test Configurations with baseline HUA32, AWS16, AZURE4} 
\centering
\label{tab:table5}
\begin{adjustbox}{width=1\textwidth}
\begin{tabular}{|c|c|c|c|c|c|c|c|c|c|c|}
\hline
\multicolumn{2}{|c|}{Testbed} & \multicolumn{9}{c|}{Set Rate (msg/min)} \\ \hline
Testbed & Max Worker &\multicolumn{3}{c|}{12} & \multicolumn{3}{c|}{30} & \multicolumn{3}{c|}{60} \\ \hline 
&  Containers & \multicolumn{1}{c|}{Execution Time} & \multicolumn{1}{c|}{Waiting Time} & \multicolumn{1}{c|}{Total Time} & \multicolumn{1}{c|}{Execution Time} & \multicolumn{1}{c|}{Waiting Time} & \multicolumn{1}{c|}{Total Time} & \multicolumn{1}{c|}{Execution Time} & \multicolumn{1}{c|}{Waiting Time} & \multicolumn{1}{c|}{Total Time} \\ \hline 
\multirow{3}{*}{HUA} & 32 & 0.00\% & 0.00\% & 0.00\% & 8.87\% & 14.77\% & 9.03\% & 1008.89\% & 69946.56\% & 2868.85\% \\ \cline{2-11}
& 16 & \multicolumn{1}{c|}{ -0.50\%} & \multicolumn{1}{c|}{8.63\%} & \multicolumn{1}{c|}{ -0.26\%} & \multicolumn{1}{c|}{7.87\% }& \multicolumn{1}{c|}{25.16\%} & \multicolumn{1}{c|}{8.34\%} & \multicolumn{1}{c|}{394.67\%} & \multicolumn{1}{c|}{78205.41\%} & \multicolumn{1}{c|}{2494.02\%}\\ \cline{2-11}
& 8 & \multicolumn{1}{c|}{-2.35\%} & \multicolumn{1}{c|}{8.12\%} & \multicolumn{1}{c|}{-2.07\%} & \multicolumn{1}{c|}{8.78\%}& \multicolumn{1}{c|}{32.83\%} & \multicolumn{1}{c|}{9.43\%} & \multicolumn{1}{c|}{103.48\%} & \multicolumn{1}{c|}{73593.77\%} & \multicolumn{1}{c|}{2086.26\%}\\ \hline

\multirow{3}{*}{AWS} & 16 & -56.45\% & -74.02\% & -56.93\% & -45.45\% & -92.54\% & -46.72\% & 122.24\% & 10746.06\% &  408.88\% \\ \cline{2-11}
& 8 & -55.38\% & -73.08\% & -55.86\% & -44.47\% & -78.12\% & -45.37\%& -0.25\% & 6486.98\% & 174.77\% \\ \hline
\cline{2-11}
\multirow{3}{*}{AZURE} & 4 & -61.30\% & -94.44\% & -62.20\% & -59.50\% & -96.01\% & -60.49\% & -57.79\% & 13275.80\% & 301.95\% \\ \cline{2-11}
& 2 & -60.07\% & -95.31\% & -61.02\% & -59.94\% & 19964.47\% & 480.32\% & -59.93\% & 31940.89\% & 803.46\% \\ \hline
\end{tabular}
\end{adjustbox}

\end{table}

\noindent  Auto-scaling techniques have been widely studied to manage dynamic container allocation in Kubernetes. Threshold-based auto-scaling techniques have been used to adjust container numbers based on thresholds of CPU or Memory usage rates \cite{altaf2018auto}. Feedback control methods, such as linear-performance-model-based fixed gain \cite{lu2006feedback, pan2008feedback}, adaptive \cite{cai2021inverse}, or multi-model switching feedback control methods \cite{patikirikorala2011multi}, have been widely used in application resource auto-scaling \cite{hellerstein2004feedback}. Such methods should be applied in order to regulate the maximum container slots dynamically so that the function execution time (or its distribution) is relatively similar over time, given the availability of relevant monitoring data\cite{physicsDE}.\par

\noindent However, in the case of FaaS platforms, and given  the availability of monitoring information across system stages , e.g. wait time, initialization time, execution time etc., more fine grained elasticity may be applied. For example, if monitoring indicates a rise in execution time, this might mean reducing the memory setting in order to reduce the number of concurrent containers. If, on the other hand, we have such a configuration and we observe high waiting times, it is an indication that further compute nodes should be added to the kubernetes cluster in a horizontal scaling manner or the used nodes can be increased mainly from a number of available processors point of view in a vertical scaling manner. \par

\subsection{Transient Phenomena Stabilization}

\noindent In many cases, knowing the time it takes for the system to stabilize can be beneficial. For example, in some modelling cases such as PID (Proportional– \\ Integral– Derivative) controller-based regulation of resources \cite{singh2019research}, a temporarily increased  derivative of the execution times may lead to corrective actions such as cluster resizing. In some cases however, we may first need to see the steady state before actually deciding on the corrective action since the steady state may still be within the set goals of the system. In that case we could avoid oscillation of the system metrics caused by premature corrective actions. Thus such information that can be extracted for example from Fig. 2g (the time it takes to reach a more steady condition in the execution time) can aid in adding such safeguards by adding a relevant lag in the autoscaling mechanisms.\par

\noindent From the graphs of section 3 with relation to wait time, the shape and form of the according graphs is very similar to a type of sigmoidal function. Thus, a relevant neural network architecture based on such a transfer function can easily depict the waiting time transition, if the input is the elapsed time from a step-wise burst such as the load applied in these experiments.  \par

\section{Conclusion}
\noindent This paper presents extensive experimentation on a serverless Function as a Service (FaaS) system, yielding empirical insights into its performance characteristics and optimization opportunities. Through diverse experiments across cloud/ edge locations and varying configurations, we gained understanding of transient effects on waiting and execution time, identifying trade-offs in system setup, and analyzed the probability of failures due to image pulling bottlenecks. The findings emphasize the need to consider cold starts not only for extended initialization times, but also for higher failure probabilities. Furthermore, the analysis highlights the importance of effective container management and execution strategies to enhance system performance and user experience. \par
\noindent Overall, the work contributes to the broader understanding of FaaS system and informs their design and optimization parameters. Furthermore, we explore trade-offs associated with system setup, which shed light on the preference for either minimizing wait time or execution time. In the long term, the empirical insights of this study can support and inspire future researchers and practitioners that design and apply FaaS systems. The explored trade-offs depending on the system scenarios may inform them towards implementing optimized systems with satisfied QoS. \par

\section{ACKNOWLEDGMENT}

\noindent The research presented has received funding from the European
Union’s Project H2020 PHYSICS (GA 101017047).\par

\bibliographystyle{elsarticle-num}
\bibliography{reference.bib}



\end{document}